# Shot-noise-limited few-cycle mid-infrared frequency comb with attosecond phase stability

Maciej Kowalczyk[1*], Jakub Jaworski[1], Michał Pietrzak[1], Karolina Suliga[1], Paweł Szczypkowski[2], Paweł Kaczmarek[1], Alexander Weigel[3,4], Jarosław Sotor[1]

[1]*Faculty of Electronics, Photonics and Microsystems, Wrocław University of Science and Technology, Wybrzeże S. Wyspiańskiego 27, 50-370 Wrocław, Poland*
[2]*Faculty of Physics, University of Warsaw, Pasteura 5, 02-093 Warsaw, Poland*
[3]*Center for Molecular Fingerprinting Research, 1093 Budapest, Czuczor Utca 2-10, Hungary*
[4]*Max-Planck-Institute of Quantum Optics, Hans-Kopfermann-Str. 1, 85748 Garching, Germany*

*Corresponding author email: maciej.kowalczyk@pwr.edu.pl

## Abstract

Achieving sub-cycle waveform control with attosecond-level precision while reaching shot-noise-limited amplitude stability in the mid-infrared spectral range remains a central challenge for ultrafast and precision optical science. Here, we demonstrate a fully stabilized, ultra-stable optical frequency comb (OFC) based on a Kerr-lens mode-locked Cr:ZnS laser operating at 2.3 µm. The oscillator delivers 40 fs pulses at a 25 MHz repetition rate, which are spectrally broadened to cover two octaves and compressed to 1.4-cycle 11 fs duration. Pumped by a custom low-noise erbium-doped fiber amplifier, the laser exhibits an integrated relative intensity noise (RIN) of 0.0026% over 10 Hz – 1 MHz, while the noise of the few-cycle output is further suppressed to 0.0017% and operates at the shot-noise limit for Fourier frequencies above 5 kHz. It represents the lowest amplitude noise reported to date for any mode-locked laser. This unprecedented amplitude stability enables carrier-envelope phase (CEP) stabilization with a residual integrated phase noise of only 1.5 mrad (10 Hz – 12.5 MHz), corresponding to a CEP jitter of 1.8 as – the highest phase stability for any laser system ever reported. The long-term performance of the fully-stabilized OFC is measured over 24 hours with a power stability of 0.01% and a residual CEP noise of 17 mrad. By combining few-cycle mid-infrared pulses with shot-noise-limited intensity noise and attosecond-level phase stability, the reported OFC provides access to new regimes of quantum-limited metrology and control of ultrafast light-matter interactions.

## Introduction

Optical frequency combs (OFCs) provide a phase-coherent link between optical and microwave domains[1,2], forming a precise frequency ruler that has transformed optical science. Over the past two decades, this technology has enabled transformative advances in optical metrology[3,4], attosecond science[5,6] and precise spectroscopy[7]. Extending OFCs into the mid-infrared (mid-IR, 2–20 µm)[8] is of central importance, as this spectral region hosts fundamental molecular ro-vibrational transitions, enabling highly sensitive and selective detection of chemical species for applications ranging from environmental monitoring[9], to biomedical diagnostics[10,11], while also benefiting strong-field physics[12,13].

However, the performance of OFC-based techniques is governed not only by spectral coverage, but also by amplitude and phase stability. Residual relative intensity noise (RIN) directly limits the achievable signal-to-noise ratio (SNR) and prevents averaging toward the quantum-noise-limited regime, thereby setting a fundamental constraint on precision spectroscopy[14]. Reaching the ultimate shot-noise limit is therefore a critical benchmark: it indicates that classical amplitude noise has been suppressed to the point where the measurement sensitivity is governed by intrinsic photon statistics rather than technical fluctuations. This is especially important in the context of emerging quantum-enhanced spectroscopy, where recent demonstrations have shown that excess RIN determines whether quantum advantage can be fully exploited in practice[15,16]. In parallel, excess laser noise has also been identified as a limiting factor for high-fidelity quantum control in atomic and qubit systems[17].

At the same time, fluctuations of the carrier-envelope phase (CEP) impose intrinsic constraints on the temporal stability of the optical waveform, translating into timing jitter of the electric field and hindering access to sub-cycle dynamics. Excess jitter may prevent long-term coherent averaging in dual-comb measurements[18], and ultimately limit the achievable sensitivity in OFC-based timing transfer as it approaches the quantum-noise-limited regime[19]. Moreover, CEP fluctuations degrade the performance of the field-resolved spectroscopy techniques[20], and limit the achievable degree of control over ultrafast light–matter interactions. Achieving improved waveform stability is therefore of fundamental importance for emerging strong-field applications such as lightwave-driven petahertz electronics[21,22], and for exploring the quantum nature of high-harmonic generation[23].

These considerations establish amplitude and phase noise not merely as technical imperfections, but as fundamental parameters governing access to measurement regimes that remain largely inaccessible today. Consequently, few-cycle mid-infrared OFCs combining shot-noise-limited intensity noise with attosecond-level phase stability are a key prerequisite for quantum-limited and quantum-enhanced detection, as well as field-resolved control of light–matter interactions.

Although significant progress has been achieved in mature near-IR OFCs, with a residual phase noise on the order of ~10-20 mrad corresponding to a jitter as low as 6 as[24–29], these advances have not yet been fully extended to the mid-IR. While mid-IR OFC has been reported in a thulium fiber system[30], its performance remained limited by relatively high residual phase noise exceeding 300 mrad. Direct longer-wavelength sources such as fluoride fiber lasers[31] and Fe:ZnSe oscillators[32] lack full OFC stabilization. Hence, indirect approaches based on nonlinear conversion of near-IR lasers are commonly employed[33,34]. They offer passive phase stability but suffer from low conversion efficiencies, typically below 1%[8], restraining their applicability.

Cr:ZnS/Se lasers emitting around 2.3 μm represent a promising platform for mid-IR OFC generation. Owing to their broad gain bandwidth[35,36], these systems support single-cycle pulse generation[37] and record-efficient (>10%) nonlinear down-conversion to longer wavelengths[38,39]. However, previously demonstrated Cr:ZnS/Se-based OFCs have been limited in terms of amplitude and phase noise, with reported RIN above 0.05%[40–42] and residual phase noise on the order of 75 mrad[43]. Our recent advances have improved these figures to 0.036% and 5.9 mrad[37], representing the highest stability achieved for few-cycle mid-IR sources prior to this work, yet without full OFC stabilization.

Here, we present a fully-stabilized ultra-low-noise OFC based on a Cr:ZnS mode-locked laser operating at 2.3 μm. The laser produces 40 fs pulses at 25 MHz repetition rate, which are spectrally broadened to two-octave spanning bandwidth and compressed to 11 fs, equivalent to 1.4 optical cycles. By employing a custom low-noise erbium-doped fiber amplifier pump source, we achieve ultrastable operation of the Cr:ZnS laser with RIN amounting to 0.0026% (integrated over the 10 Hz – 1 MHz range). Nonlinear spectral-broadening provides additional passive noise suppression, reducing the RIN down to the fundamental shot-noise limit for Fourier frequencies above 5 kHz in the few-cycle regime (0.0017%). This constitutes over an order-of-magnitude improvement over state-of-the-art Cr:ZnS/Se systems[37] and represents the most amplitude-stable mode-locked laser ever reported[44]. Operation at the shot-noise limit confirms effective suppression of classical intensity noise and provides direct access to quantum-noise-limited performance. This unprecedented amplitude stability enables carrier-envelope phase (CEP) locking with a residual integrated phase noise of 1.5 mrad (10 Hz – 12.5 MHz), corresponding to a phase jitter of 1.8 as, and represents the highest carrier-envelope offset frequency ($f_{ceo}$)-stability reported for any laser to date. Full OFC stabilization is achieved by locking the repetition rate frequency ($f_{rep}$) using balanced optical–microwave phase detection (BOMPD) technique. Finally, we demonstrate long-term stability over 24 hours, reaching an average power standard deviation as low as 0.01%, together with a residual CEP noise of 17 mrad. Together, these results establish a new operating regime for mid-IR frequency combs, where few-cycle pulse duration, shot-noise-limited intensity noise, and attosecond-level phase stability are achieved simultaneously in a fully stabilized platform for quantum-limited and field-resolved measurement science.

## Results

### Mid-IR frequency comb architecture

The amplitude stability of any laser oscillator is mainly limited by the fluctuations of its pump source. Moreover, pump fluctuations also influence the free-running $f_{ceo}$ linewidth, and thus limit the ultimate phase stability, which can be achieved in a stabilized laser[45]. Typically, mode-locked Cr:ZnS/Se lasers are pumped by erbium-doped fiber lasers characterized by strong relaxation oscillations, which are further transferred to the laser noise, resulting in integrated RIN above 0.1%[36,42,46,47]. In our previous work, we have shown that direct diode pumping allows to decrease the RIN down to 0.036% (the best result achieved for Cr:ZnS/Se lasers prior to this work)[37]. Here, we employed an alternative approach that boosts the power of a low-noise, narrow-linewidth laser-diode seed in an erbium-doped fiber amplifier (EDFA). We developed an ultrastable EDFA system based on a classic master-oscillator power-amplifier architecture. The EDFA laser system delivers up to 6 W of linearly polarized output power at 1560 nm with a polarization extinction ratio exceeding 27 dB. The EDFA output power can be either absolutely stabilized or locked to an external feedback voltage signal, enabling active stabilization of the pumped laser system output. Experimental details of the EDFA pump laser can be found in Supplementary Materials, section S1.

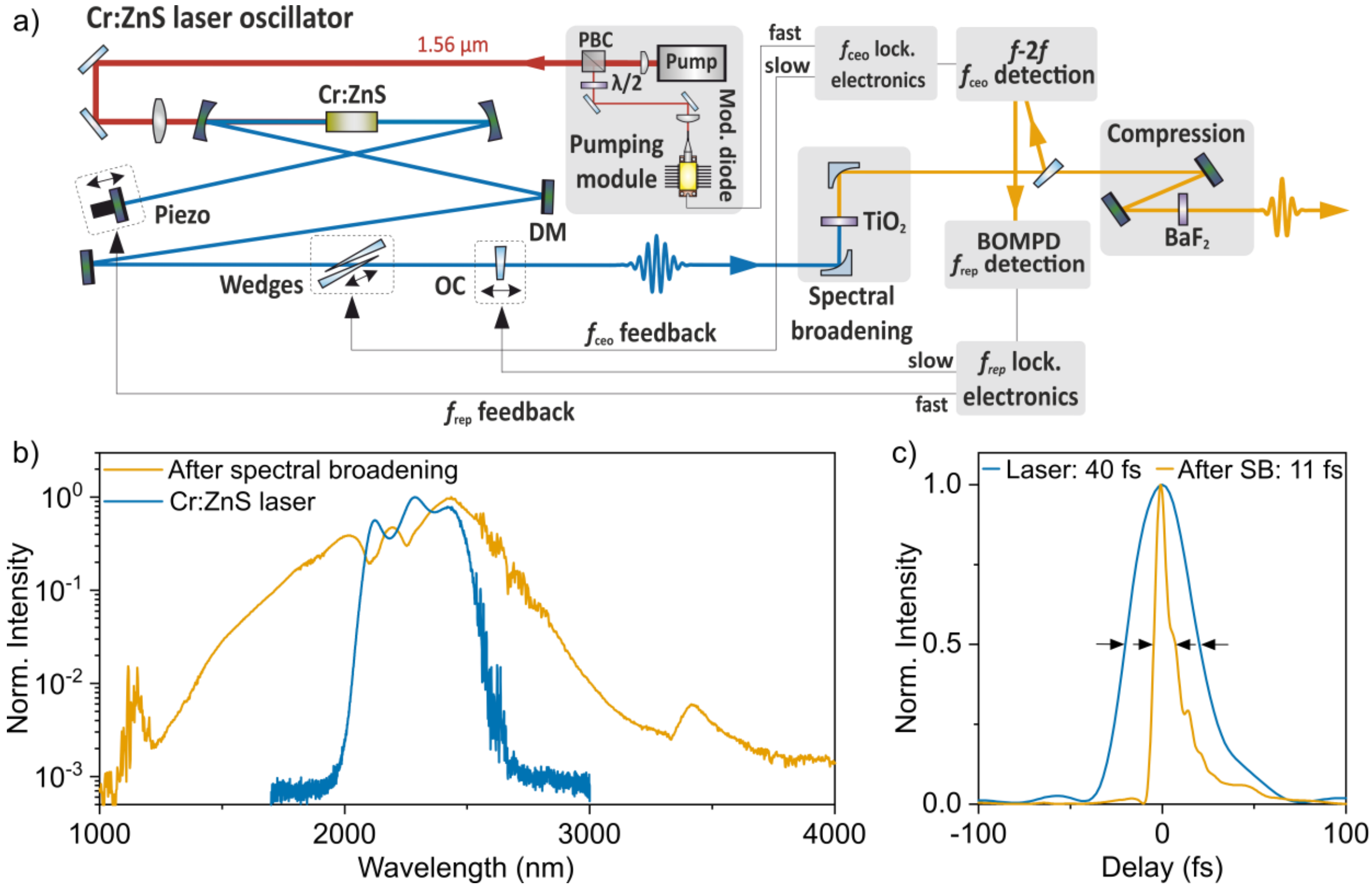

**Fig. 1. Mid-IR optical frequency comb. (a)** Experimental setup of the Cr:ZnS frequency comb: the laser output is spectrally broadened in a $TiO_2$ plate and incident on a beamsplitter: main part of the beam is compressed by chirped mirrors, while the reflected beam is used for frequency comb stabilization. Actuators for fast and slow $f_{rep}$ and $f_{ceo}$ stabilization loops are marked. **(b)** Optical spectra of the beam directly emitted from the Cr:ZnS laser oscillator (blue); and after spectral broadening in a $TiO_2$ plate (orange); **(c)** Corresponding pulse profiles retrieved from SHG-FROG.

Figure 1a presents a simplified schematic of our frequency comb based on a soft-aperture Kerr-lens mode-locked Cr:ZnS laser. It is pumped by ~4.2 W from the EDFA pump source. The laser produces an average power of 515 mW at 25 MHz, resulting in a pulse energy of 21 nJ. The spectrum is centered at 2.3 µm, and covers 2 – 2.7 µm range at −20 dB scale (Fig. 1b). The pulse duration was measured by a home-built second harmonic generation frequency-resolved optical gating (SHG-FROG) setup and amounts to 40 fs (Fig. 1c). Experimental details of the Cr:ZnS laser can be found in Methods section.

The output of the Cr:ZnS laser oscillator is spectrally broadened in a 500-µm-thick crystalline ([001]-cut) $TiO_2$ plate[37]. The output spectrum spans over two optical octaves from 1 to 4 µm (Fig. 1b). The peaks around 1.15 µm originate from a residual SHG signal generated in the oscillator, due to random quasi-phase-matching in Cr:ZnS gain medium[36]. The average power after the plate amounts to 400 mW, with a 16% loss occurring due to light absorption by plasma generated in the plate[37]. After spectral broadening, the beam is sent onto a beam splitter, consisting of a thin ZnS window, with one side AR-coated. Due to Fresnel reflection, ~15% (~60 mW) of the signal is backreflected and used for $f_{ceo}$ detection in a $f$-2$f$ interferometer, while the transmitted beam (~85%, ~340 mW) is available for further experiments. This beam is subsequently compressed by a combination of a single pair of complementary dispersive mirrors providing second- and third-order dispersion[48] with an additional fine-tuning provided by bulk $BaF_2$. The temporal profile of the pulses was measured by a SHG-FROG setup based on Si and InGaAs spectrometers, resulting in 11 fs pulse duration (Fig. 1c), equivalent to 1.4 optical cycle.

For OFC stabilization, $f_{ceo}$ is locked via pump-power modulation using an additional pump diode (fast actuator). It is assisted by a slow correction from a motorized intracavity wedge pair, when the available range of the modulation diode is exceeded. $f_{rep}$ stabilization is realized via a piezoelectric transducer controlling resonator's end mirror position (fast actuator), combined with slow correction of the output coupler (OC) position placed on a motorized stage.

## Relative intensity noise characterization

First, we characterized the RIN of the pump and the mode-locked Cr:ZnS lasers by using the standard procedure described in Supplementary Materials, section S2. For all RIN measurements the noise was measured over the range 10 Hz – 1 MHz frequency range, with the upper limit set by the photodetector bandwidth. Above 1 MHz, the laser is expected to operate at the shot-noise limit, due to the ~4 µs fluorescence lifetime[49] of the Cr:ZnS gain medium (see CEP measurements in Fig. 3b). Figure 2a presents the power spectral density (PSD) of the RIN and the corresponding integrated noise for the free-running Cr:ZnS laser with an unprecedentedly low RIN of 0.0026%. The improvement exceeding an order of magnitude over our previous report (0.036%)[37], was achieved primarily by replacing diode pumping with our ultra-stable EDFA pump source. RIN data for the EDFA pump laser is also shown in Fig. 2a with a corresponding integrated RIN of 0.0037%. The noise of the mode-locked Cr:ZnS laser is lower than that of the pump at frequencies above ~100 kHz, due to low-pass filtering by the Cr:ZnS crystal of fluctuations faster than the gain medium's response time. At lower frequencies, the noise curves overlap, confirming that the RIN of the laser oscillator is limited by the pump noise.

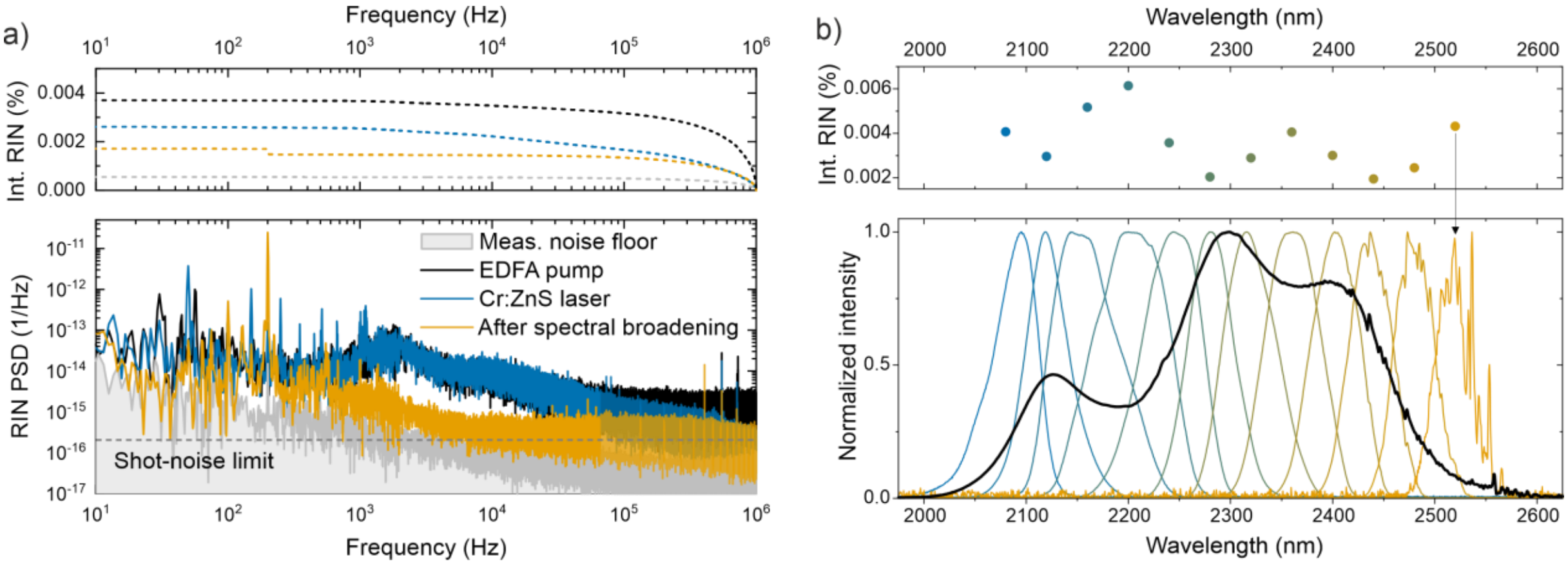

**Fig. 2. Relative intensity noise of the Cr:ZnS laser. (a)** RIN PSD (solid lines, lower panel) and the corresponding integrated RIN (dashed line, upper panel) for the measurement noise floor limit (grey; RIN = 0.0005%), EDFA pump laser (black; 0.0037%); direct Cr:ZnS laser output (blue; 0.0026%), and after spectral broadening (orange; 0.0017%). Dashed horizontal line shows the shot-noise RIN limit (the same optical power was incident on the detector for all measurements). **(b)** Spectrally-resolved RIN of the free-

running mode-locked Cr:ZnS laser: the isolated spectral components are shown with the entire spectrum in the lower panel, while the upper panel shows the corresponding integrated RIN.

To gain deeper insight into the spectral nature of the residual intensity noise, we further investigated how the RIN varies across the laser emission bandwidth. We studied the spectral dependence of the laser intensity noise, as described in the Methods section. Figure 2b presents the normalized spectra compared with the laser output and the corresponding integrated RIN values. The RIN is clearly not constant, but, on the contrary, it is anticorrelated with the spectral intensity. It spans from 0.006% (at a local minimum of spectral intensity at 2200 nm), down to 0.002% (at local maxima at 2280 nm and 2440 nm). Interestingly, in the region above ~2450 nm where several spectral peaks and dips are present due to interaction of the intracavity light with water vapor[50], no excessive increase of the RIN is observed. Our result confirms that our ultra-low-noise performance of the laser output is not concentrated on small spectral islands, but is characteristic over the full output bandwidth of the oscillator. Moreover, our study anticipates that spectral shaping of the Cr:ZnS output to achieve more homogenous spectra can be a viable path towards further reducing the laser RIN below 0.0026% achieved in this work.

Subsequently, we measured the RIN of the spectrally broadened beam after the $TiO_2$ plate. Figure 2a compares it with the noise of the direct Cr:ZnS laser output. The highly nonlinear nature of spectral broadening in $TiO_2$ including excitation of free carriers[37], gives rise to intensity-dependent (plasma-related) losses that clamp intensity fluctuations by a saturation-like response[51]. This mechanism provides a fully passive amplitude noise suppression, resulting in an integrated RIN of only 0.0017% for sub-two-cycle 11 fs pulses. Remarkably, the RIN is suppressed down to the fundamental shot-noise limit at frequencies higher than ~5 kHz. We confirm that the signal is shot-noise limited for the power levels up to the saturation level of the employed photodetector (see Supplementary Materials, section S2). To the best of our knowledge, this result constitutes the lowest RIN for any mode-locked laser ever reported[44].

## OFC stabilization

Access to super-octave-spanning spectrum enables $f_{ceo}$ detection in an $f$-2$f$ interferometer. The registered RF spectrum is shown in Fig. 3a. The SNR of the detected $f_{ceo}$ beat-notes at a 100 kHz resolution bandwidth (RBW) is ~75 dB, with an absolute RF power reaching 0 dBm. The signal strength is limited by the saturation of the employed detector, resulting in the appearance of intermodulation peaks generated in the detector, which, however, do not deteriorate the lock. As shot-noise is an ultimate factor limiting achievable phase stability[28,52], a high $f_{ceo}$ SNR is essential for achieving ultra-low-noise operation.

The inset in Fig. 3a shows a zoom-in of the $f_{ceo}$ beat-note measured at 200 kHz frequency span measured within a time window of 38 ms (RBW = 100 Hz). The $f_{ceo}$ frequency is directly correlated with the intracavity power, and hence fluctuations of the latter cause broadening of the linewidth[45]. The exceptionally narrow linewidth of the free-running $f_{ceo}$ signal is enabled by the ultrastable performance of the Cr:ZnS laser, supported by the low-noise EDFA pump source. We studied how the $f_{ceo}$ linewidth depends on the RIN of the Cr:ZnS laser. For this, we compared pumping with our EDFA source to a commercial pump laser, which is not RIN-optimized, and found that this led to roughly 10 times higher RIN of the Cr:ZnS laser resulting in a ~500 kHz $f_{ceo}$ linewidth. More details can be found in the Supplementary Materials, section S3.

Here, we employed absolute CEP-stabilization (i.e. $f_{ceo} = 0$) by using a self-referenced locking scheme based on locking the $f_{rep} \pm f_{ceo}$ signals to $f_{rep}$[37,53]. However, the $f_{ceo}$ can also be locked to any arbitrary frequency using a standard stabilization scheme that employs an external RF reference signal. All locking electronics, including a modulation diode driver, were developed in-house and customized to achieve the lowest possible residual phase noise. For $f_{rep}$ stabilization we employed a balanced optical–microwave phase detection (BOMPD) technique[54]. Experimental details of the $f_{ceo}$ and $f_{rep}$ stabilization schemes can be found in Supplementary Materials, sections S4 and S5, respectively.

## OFC characterization

The residual $f_{ceo}$ noise was measured in a separate out-of-loop (OOL) $f$-2$f$ interferometer using a partially-attenuated beam transmitted through a ZnS beam-splitter (see Methods section for details). Figure 3b presents the phase noise PSD and the corresponding integrated noise measured for the locked laser in the frequency range from 10 Hz to the Nyquist frequency ($f_{rep}$/2) of 12.5 MHz. Two servo bumps are present: the first at ~200 kHz can be associated with the PID integrator component cut-off, while the second at ~600 kHz originates from low pass-filtering due to gain-medium dynamics[55]. Above ~1 MHz we reach the fundamental phase noise shot noise level ($1.4\cdot10^{-13}$ $rad^2$/Hz) estimated for the detected $f_{ceo}$ beat-note SNR (Fig. 3a) based on the methodology adapted from ref.[52]. The phase noise shot-noise limit integrated in the entire available range of Fourier frequencies is 0.9 mrad. The residual integrated phase noise of the Cr:ZnS laser is only 1.6-times higher, and amounts to 1.48 mrad, with a corresponding CEP jitter of 1.8 as (at 2.3 µm). This constitutes the highest $f_{ceo}$-stability ever achieved for any mode-locked laser[24–26,37,56]. The $f_{ceo}$ fluctuations are intrinsically correlated with the pulse energy dynamics[57], thus, the four-fold improvement of the $f_{ceo}$-stability regarding to our previous best result (5.9 mrad)[37], has been predominantly enabled by improving the Cr:ZnS laser RIN (from 0.036%) by an order of magnitude down to 0.0026%, which was realized by replacing direct diode pumping with the ultrastable EDFA pump source. A comparison of the presented results against directly diode-pumped Cr:ZnS laser from ref.[37] can be found in Supplementary Materials, section S6.

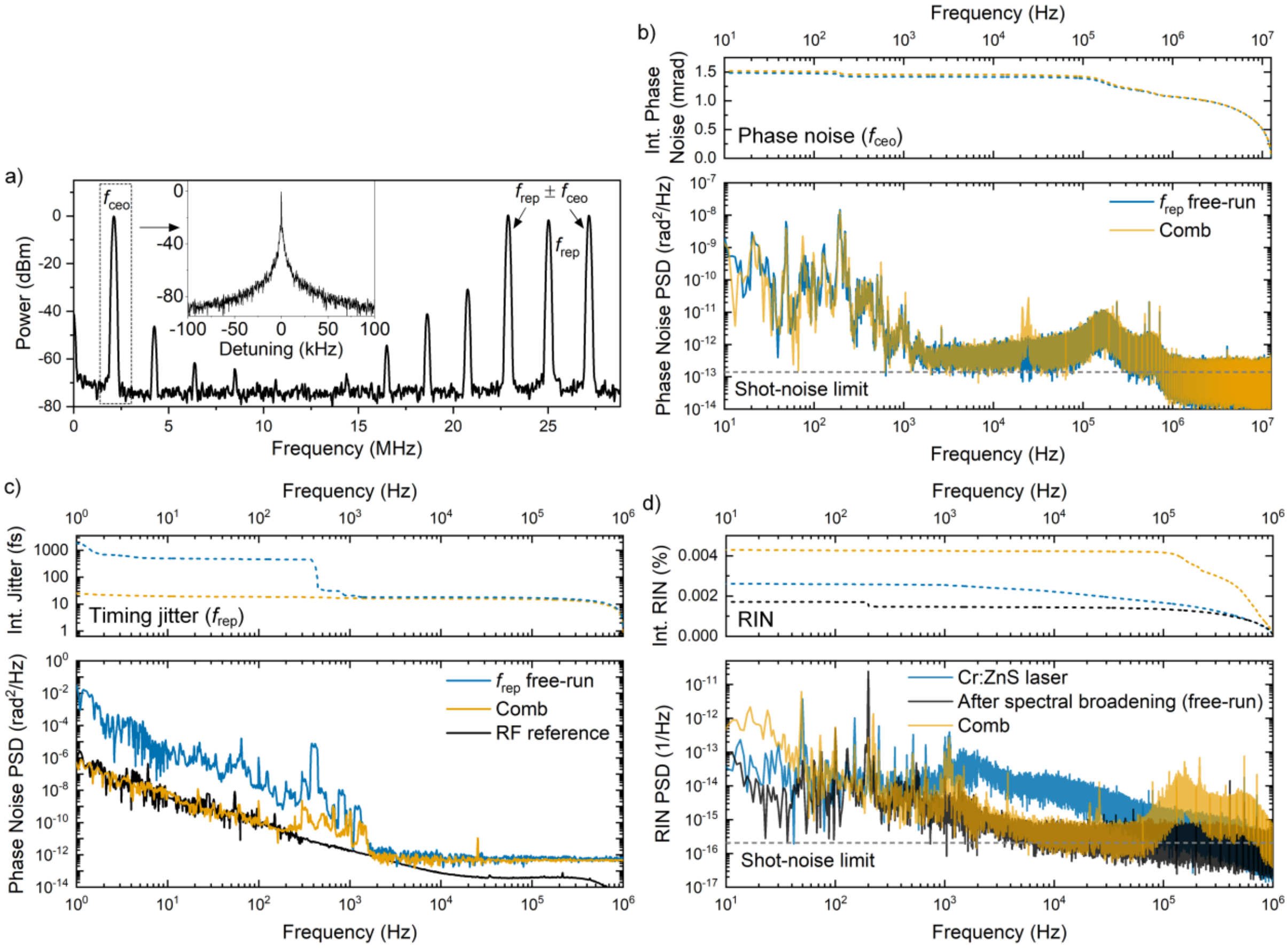

**Fig. 3. Mid-IR optical frequency comb performance. (a)** RF spectrum showing the first $f_{rep}$ harmonic of the laser at 25 MHz, and the $f_{ceo}$ beat-notes (resolution bandwidth (RBW) = 100 kHz; video bandwidth (VBW) = 10 kHz). Low-power unmarked peaks correspond to intermodulation signals generated in the photodetector. Inset: zoom-in presenting the free-running $f_{ceo}$ beat-note (RBW = VBW = 100 Hz, 38 ms measurement time). **(b)** $f_{ceo}$ phase noise PSD (solid lines, lower panel) and corresponding integrated phase noise (dashed line, upper panel) for the CEP-locked Cr:ZnS laser with a free-running $f_{rep}$ (blue; 1.48 mrad), and for the fully-stabilized frequency comb ($f_{ceo}$ and $f_{rep}$ locked; orange; 1.5 mrad). Dashed horizontal line indicates the phase noise shot-noise limit originating from the $f_{ceo}$ SNR shown in (a). **(c)** $f_{rep}$ phase noise PSD (solid line, lower panel) and the corresponding timing jitter (dashed line, upper panel) for the free-

running (blue; 1900 fs) and $f_{rep}$-locked (orange; 25 fs) Cr:ZnS laser. Phase noise PSD of the employed RF reference source (SMA 100B; R&S) is shown in black. **(d)** RIN PSD (solid lines, lower panel) and the corresponding integrated RIN (dashed line, upper panel) measured for the direct output of the free-running Cr:ZnS laser (blue; 0.0026%), after spectral broadening (black; 0.0017%), and the fully-stabilized OFC ($f_{ceo}$ and $f_{rep}$ locked; orange; 0.0043%). Dashed horizontal line indicates the shot-noise RIN limit (the same optical power was incident on the detector for all measurements).

In the next step, we established full OFC operation by stabilizing the $f_{rep}$ frequency with the BOMPD-based feedback loop. The residual $f_{rep}$ noise PSD with the $f_{rep}$ jitter integrated in the range of 1 Hz – 1 MHz is shown in Fig. 3c. The integrated jitter amounts to 1900 fs for the free-running laser, and 25 fs for the stabilized OFC. However, this result is currently limited by the stability of the employed RF reference (SMA100B; R&S, black line) for the low frequencies below 1 kHz. At higher Fourier frequencies, the measured noise is limited by the signal-to-noise ratio of the $f_{rep}$ beatnote, which is constrained by the detector response and results in a flat noise floor. Figure 3b presents the comparison of the $f_{ceo}$ phase noise for the free-running and locked $f_{rep}$, with the latter case introducing only a minor additional noise component around 25 kHz, where the resonance of the employed PZT occurs. The integrated CEP noise for the OFC amounts to 1.5 mrad.

Finally, we investigate how frequency comb stabilization affects the RIN of the laser oscillator. Figure 3d compares the RIN PSD of the free-running Cr:ZnS laser (blue) and the full comb (orange), both measured at the direct laser output before spectral broadening stage. As intensity noise is intrinsically correlated with $f_{ceo}$ fluctuations, OFC stabilization suppresses RIN by roughly an order of magnitude in the intermediate frequency range from 500 Hz to 100 kHz. At low frequencies, mechanical noise dominates, while at high frequencies, the servo bumps introduced by the $f_{ceo}$ and $f_{rep}$ stabilization loops (cf. Fig. 3b) are transferred to the intensity noise increasing the total integrated RIN to 0.0043%. Active noise suppression enables reaching the fundamental shot-noise limit in the ~5 – 100 kHz range. Remarkably, the RIN curve for the full OFC overlaps with the noise measured after spectral broadening without stabilization. This demonstrates that the reported ultrastable Cr:ZnS laser can reach the fundamental shot-noise RIN limit with either passive or active stabilization.

## OFC long-term characterization

While the above noise characterization describes the short-term OFC stability, several applications, including high-resolution spectroscopy(*7,18*), coherent signal averaging(*14,16*), and precision metrology (*2,3*), rely on extended averaging times and thus are limited by the long-term OFC drifts. We have accordingly studied a long-term performance of our Cr:ZnS frequency comb measured over 24 hours. First, we investigated the power stability of the Cr:ZnS laser oscillator by using a standard power meter. The average power of the Cr:ZnS laser was stabilized by controlling the power of the EDFA pump laser with a slow feedback loop based on the power meter signal. Figure 4a presents the average power of the power-stabilized Cr:ZnS laser compared with the variations of the relative humidity (RH) and temperature in the laser chamber. For this measurement, the RH stabilization was not engaged, and the RH in the laser chamber followed the room conditions. The average power standard deviation amounted to 0.01%, limited by the accuracy of the employed power meter.

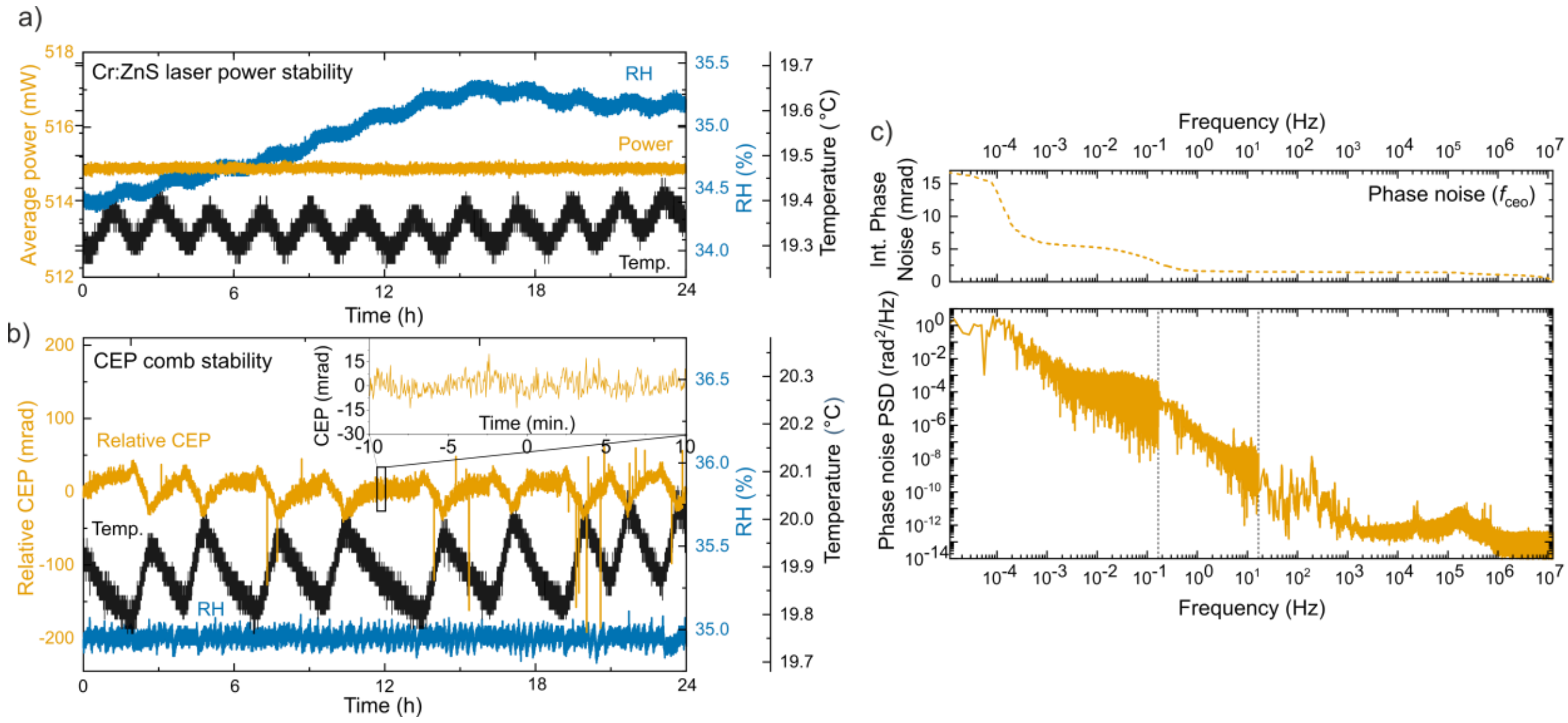

**Fig. 4. Mid-IR optical frequency comb long-term (24 h) performance. (a)** Average power stability of the Cr:ZnS laser (orange) compared with the trend of environmental conditions in the laser chamber: RH (unlocked, blue) and temperature (black). **(b)** CEP stability of the Cr:ZnS laser (orange) compared with the trend of environmental conditions in the laser chamber: RH (locked, blue) and temperature (black). The inset shows CEP fluctuations over 20 minutes. **(c)** Phase noise PSD (solid line, lower panel) and corresponding integrated phase noise (dashed line, upper panel) for merged data from short- (>10 Hz, cf. Fig. 3b) and long-term measurements (<16 Hz, one data set for 12 μHz – 0.16 Hz, cf. Fig. 4b; second data set for 0.16 Hz – 16 Hz). Frequency ranges for the three data sets are indicated by dashed vertical lines.

Subsequently, we measured long-term CEP stability in an OOL *f*-2*f* interferometer by analyzing the spectral stability of the interference fringes acquired by the spectrometer[58]. Figure 4b presents a measurement with the DC signal from the fast OOL phase noise measurement (cf. Fig. 3b) fed as a correction to the CEP-stabilization loop to prevent phase drifts. For this measurement RH was actively stabilized with a peak-to-peak deviation of ~±0.1% (see Methods). Residual phase noise amounts to 15 mrad, with the CEP fluctuations being correlated with laboratory environment conditions. In Supplementary Materials, section S7 we present a comparison of average power and CEP stability with and without additional slow-feedback correction.

Phase noise data from the 24-h-long measurement have been Fourier-transformed and merged with the PSD shown in Fig. 3b. The total CEP noise of our OFC integrated over 12 frequency decades from 12 μHz to 12.5 MHz amounts to an exceptionally low value of 17 mrad, corresponding to a phase jitter of 21 as (Fig. 4c).

# Discussion

In conclusion, we have demonstrated an ultra-stable, few-cycle mid-infrared optical frequency comb based on a Kerr-lens mode-locked Cr:ZnS 25-MHz laser operating at 2.3 μm. By combining a low-noise, custom-designed erbium-doped fiber amplifier with extreme spectral broadening and pulse compression, we achieved record-low intensity noise and attosecond-level CEP stability for 1.4-cycle, 11 fs pulses. Our system achieves a relative intensity noise of 0.0026% integrated over the 10 Hz – 1 MHz range. Subsequent nonlinear spectral broadening provides additional passive noise suppression, reducing the RIN of the few-cycle output to 0.0017% and reaching the shot-noise limit for Fourier frequencies above 5 kHz. This performance represents the lowest amplitude noise reported for any mode-locked laser to date, and confirms the effective suppression of classical intensity noise down to the intrinsic photon-noise floor. Enabled by this exceptional amplitude stability, carrier-envelope phase stabilization reaches a residual integrated phase noise of only 1.5 mrad (over the 10 Hz – 12.5 MHz ($f_{rep}$/2) range), corresponding to 1.8 as phase jitter and setting a new benchmark for phase stability in mode-locked lasers. Full OFC stabilization is completed by locking the $f_{rep}$ using a low-noise BOMPD technique to the limit of the reference oscillator stability. Beyond short-term performance, we demonstrate stable 24-hour operation with an average power fluctuations reaching 0.01%, and a residual CEP noise of 17 mrad, confirming the robustness of the system for long-duration precision measurements.

In the future, additional post-amplification[59–63] will provide the pulse energies required for strong-field experiments[12,13], while nonlinear conversion toward the long-wavelength infrared[38,39] or ultraviolet[64] can extend the ultrastable OFC operation to broader spectral regions. By simultaneously suppressing classical intensity noise to the shot-noise floor, and reducing phase jitter to the attosecond level, the demonstrated MHz-rate few-cycle light source directly addresses the key noise limitations of OFC-based measurements and provides an enabling platform for advancing precision spectroscopy[7,20], quantum-enhanced metrology[15,16] and lightwave-controlled strong-field science[13,21].

# Materials and methods

## Cr:ZnS mode-locked laser

The Cr:ZnS gain medium with a slab geometry (2×5×5 $mm^3$) is AR-coated for the 1.5 – 3 μm range and exhibits ~10% transmission at 1560 nm. It is pumped by a the custom EDFA laser (Supplementary Materials, section S1). An additional pump laser diode emitting up to 120 mW at 1.56 μm (PL-DFB-1560-200-50-A81-PA-NL, LD-PD INC) is used for $f_{ceo}$ stabilization. Its output is combined with the main pump beam via a polarization beam-combiner (CCM1-PBS25-1550, Thorlabs). The 6-m-long cavity of the Cr:ZnS laser is folded using a set of dispersive mirrors allowing to tune the net dispersion for minimum duration of the generated solitonic pulses. It is terminated on one side by a wedged OC mirror with 30% transmission, while the second end mirror is glued to a piezoelectric stack (PZT) for $f_{rep}$ stabilization. The OC is placed on a motorized stage, for a coarse $f_{rep}$ correction, when its long-term drift exceeds the PZT travel range (16 μm). In addition, a pair of thin $CaF_2$ wedges is placed in the laser cavity at Brewster's angle. They provide coarse $f_{ceo}$ control across several tens of MHz allowing correction of any $f_{ceo}$ drifts, that would otherwise surpass the available modulation range of the diode and thus enable a long-term lock. Cr:ZnS/Se lasers are sensitive to RH changes due to the broadband spectral coverage overlapping with water-vapor absorption lines above 2.5 μm (see Fig. 1b)[43,65]. Therefore, the laser is housed in a hermetic aluminum chamber, with RH stabilization provided by a membrane dehumidifier (M-3J1R, Rosahl).

## Spectrally-resolved RIN characterization

A scheme of the setup for spectrally-resolved RIN measurement shown in Fig. 5. The beam emitted from the Cr:ZnS oscillator was incident on a pair of diffraction gratings introducing spatial dispersion of the spectral components. Narrowband (FWHM = ~40 nm) signals were isolated by an adjustable mechanical slit and were subsequently analyzed by a standard RIN measurement setup. The power incident on the photodetector was normalized by a tunable ND filter to generate a constant DC voltage for all measurements.

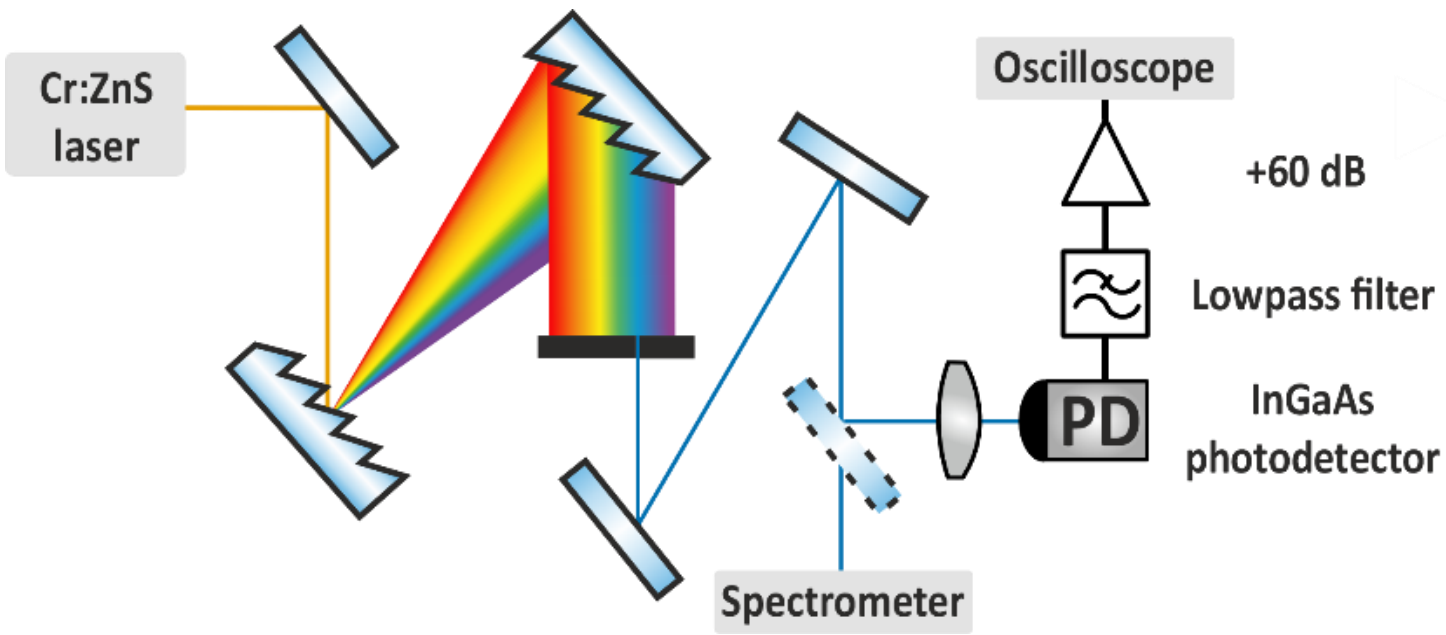

**Fig. 5.** Experimental setup for measuring the spectrally-resolved RIN.

## Optical frequency comb characterization

The residual $f_{ceo}$ noise was measured by implementing an approach presented in ref.[66] in an additional out-of-loop (OOL) $f$-$2f$ interferometer using a partially-attenuated beam transmitted through a ZnS beam-splitter. SHG was realized in a 1-mm-thick PPLN, and the signal was registered by a home-built BD, low-pass filtered at 12.5 MHz and amplified by 50 dB prior to analysis, which was performed according to the procedure described in ref.[37]. The optical signal from the OOL $f$-$2f$ is also split and sent to a spectrometer (NR-512-1.7, Ocean Optics) to quantify long-term (>16 Hz) CEP fluctuations based on the stability of the spectral fringe pattern (Supplementary Materials, section S7).

The residual $f_{rep}$ noise was quantified by measuring the monitor signal from the BOMPD setup by a fast InGaAs photodetector (DXM20AF, Thorlabs) at 380$^{th}$ harmonic (9.502 GHz) which was bandpass-filtered (ZVBP-9500-S+) and coupled to a RF phase noise analyzer (FSWP, R&S).

- **Acknowledgments**

This research was funded in whole or in part by National Science Centre (2023/02/1/ST7/00006) and Polish National Agency for Academic Exchange (BPN/PPO/2022/1/0028) within Polish Returns Programme. The project "Ultrastable pulsed lasers covering the spectral range from near to far infrared" (FENG.02.02-IP.05-0069/23) is carried out within the First Team programme of the Foundation for Polish Science co-financed by the European Union under the European Funds for Smart Economy 2021-2027 (FENG). We acknowledge the funding from the Max Planck Society within the Max Planck Partner Group Programme. This work is supported by the use of the National Laboratory for Photonics and Quantum Technologies (NPLQT) infrastructure, financed by the European Funds under the Smart Growth Operational Programme. We acknowledge Prof. Ferenc Krausz's Attoworld Group from the Ludwig Maximilian University of Munich for infrastructure support regarding the Cr:ZnS laser.

- **Author Details**

**Faculty of Electronics, Photonics and Microsystems, Wrocław University of Science and Technology, Wybrzeże S. Wyspiańskiego 27, 50-370 Wrocław, Poland**

Maciej Kowalczyk, Jakub Jaworski, Karolina Suliga, Michał Pietrzak, Paweł Kaczmarek[1], Jarosław Sotor

**Faculty of Physics, University of Warsaw, Pasteura 5, Warsaw 02-093, Poland**

Paweł Szczypkowski

**Center for Molecular Fingerprinting Research, 1093 Budapest, Czuczor Utca 2-10, Hungary**

Alexander Weigel

**Max-Planck-Institute of Quantum Optics, Hans-Kopfermann-Str. 1, 85748 Garching, Germany**

Alexander Weigel

- **Author Contributions**

The research was initiated by A.W. and led by M.K. P.K. and J.S. developed the EDFA pump source. J.S. developed the BOMPD setup. M.K., J.J., K.S., M.P., and P.S. developed the Cr:ZnS frequency comb and performed its characterization together with J.S. M.K. analyzed the data and wrote the manuscript, which was reviewed by all authors.

- **Data Availability**

The data that support the findings of this study are available from the corresponding author upon reasonable request.

- **Conflict of Interest**

J.S. runs Mode-locked Technology Sp. z o. o., which developed the low-noise EDFA pump laser and the $f_{ceo}$ locking electronics. The authors declare no other competing interests.

- **Supplementary information**

Supplementary information accompanies the manuscript on the Light: Science & Applications website (http://www.nature.com/lsa).

## References

1. Jones, D. J. *et al.* Carrier-Envelope Phase Control of Femtosecond Mode-Locked Lasers and Direct Optical Frequency Synthesis. *Science* **288**, 635–639 (2000).
2. Apolonski, A. *et al.* Controlling the Phase Evolution of Few-Cycle Light Pulses. *Phys. Rev. Lett.* **85**, 740–743 (2000).
3. Udem, T., Holzwarth, R. & Hänsch, T. W. Optical frequency metrology. *Nature* **416**, 233–237 (2002).
4. Ludlow, A. D., Boyd, M. M., Ye, J., Peik, E. & Schmidt, P. O. Optical atomic clocks. *Rev. Mod. Phys.* **87**, 637–701 (2015).
5. Hentschel, M. *et al.* Attosecond metrology. *Nature* **414**, 509–513 (2001).
6. Krausz, F. & Ivanov, M. Attosecond physics. *Rev. Mod. Phys.* **81**, 163–234 (2009).
7. Picqué, N. & Hänsch, T. W. Frequency comb spectroscopy. *Nat. Photonics* **13**, 146–157 (2019).
8. Schliesser, A., Picqué, N. & Hänsch, T. W. Mid-infrared frequency combs. *Nat. Photonics* **6**, 440–449 (2012).
9. Rieker, G. B. *et al.* Frequency-comb-based remote sensing of greenhouse gases over kilometer air paths. *Optica* **1**, 290–298 (2014).
10. Liang, Q. *et al.* Breath analysis by ultra-sensitive broadband laser spectroscopy detects SARS-CoV-2 infection. *J. Breath Res.* **17**, 036001 (2023).
11. Kepesidis, K. V. *et al.* Electric-Field Molecular Fingerprinting to Probe Cancer. *ACS Cent. Sci.* **11**, 560–573 (2025).
12. Wolter, B. *et al.* Strong-Field Physics with Mid-IR Fields. *Phys. Rev. X* **5**, 021034 (2015).
13. Hu, X., Mak, K. F., Zhang, J., Wei, Z. & Krausz, F. Ultrafast lasers for attosecond science. *Light Sci. Appl.* **15**, 24 (2026).
14. Zhong, W. *et al.* Broadband photon-counting dual-comb spectroscopy with attowatt sensitivity over turbulent optical paths. *Light Sci. Appl.* **14**, 293 (2025).
15. Herman, D. I. *et al.* Squeezed dual-comb spectroscopy. *Science* **387**, 653–658 (2025).
16. Wan, Z., Chen, Y., Zhang, X., Yan, M. & Zeng, H. Quantum correlation-enhanced dual-comb spectroscopy. *Light Sci. Appl.* **14**, 257 (2025).
17. Day, M. L., Low, P. J., White, B., Islam, R. & Senko, C. Limits on atomic qubit control from laser noise. *Npj Quantum Inf.* **8**, 72 (2022).
18. Coddington, I., Newbury, N. & Swann, W. Dual-comb spectroscopy. *Optica* **3**, 414–426 (2016).
19. Caldwell, E. D. *et al.* Quantum-limited optical time transfer for future geosynchronous links. *Nature* **618**, 721–726 (2023).
20. Pupeza, I. *et al.* Field-resolved infrared spectroscopy of biological systems. *Nature* **577**, 52–59 (2020).
21. Heide, C., Keathley, P. D. & Kling, M. F. Petahertz electronics. *Nat. Rev. Phys.* **6**, 648–662 (2024).
22. Hanus, V. *et al.* Light-field-driven current control in solids with pJ-level laser pulses at 80 MHz repetition rate. *Optica* **8**, 570–576 (2021).
23. Lemieux, S. *et al.* Photon bunching in high-harmonic emission controlled by quantum light. *Nat. Photonics* **19**, 767–771 (2025).
24. Fuji, T. *et al.* Attosecond control of optical waveforms. *New J. Phys.* **7**, 116 (2005).
25. Borchers, B., Koke, S., Husakou, A., Herrmann, J. & Steinmeyer, G. Carrier-envelope phase stabilization with sub-10 as residual timing jitter. *Opt. Lett.* **36**, 4146–4148 (2011).
26. Liao, R. *et al.* Active f-to-2f interferometer for record-low jitter carrier-envelope phase locking. *Opt. Lett.* **44**, 1060–1063 (2019).
27. Fehrenbacher, D. *et al.* Free-running performance and full control of a passively phase-stable Er:fiber frequency comb. *Optica* **2**, 917–923 (2015).
28. Endo, M., Shoji, T. D. & Schibli, T. R. Ultralow Noise Optical Frequency Combs. *IEEE J. Sel. Top. Quantum Electron.* **24**, 1–13 (2018).
29. Schmid, F. *et al.* An ultra-stable high-power optical frequency comb. *APL Photonics* **9**, 026105 (2024).
30. Gaida, C. *et al.* High-power frequency comb at 2 μm wavelength emitted by a Tm-doped fiber laser system. *Opt. Lett.* **43**, 5178–5181 (2018).

31. Zhang, Y. *et al.* Advances and Challenges of Ultrafast Fiber Lasers in 2–4 µm Mid-Infrared Spectral Regions. *Laser Photonics Rev.* **18**, 2300786 (2024).
32. Pushkin, A. V., Migal, E. A., Tokita, S., Korostelin, Y. V. & Potemkin, F. V. Femtosecond graphene mode-locked Fe:ZnSe laser at 4.4 µm. *Opt. Lett.* **45**, 738–741 (2020).
33. Hoghooghi, N. *et al.* Broadband 1-GHz mid-infrared frequency comb. *Light Sci. Appl.* **11**, 264 (2022).
34. Muraviev, A. V., Smolski, V. O., Loparo, Z. E. & Vodopyanov, K. L. Massively parallel sensing of trace molecules and their isotopologues with broadband subharmonic mid-infrared frequency combs. *Nat. Photonics* **12**, 209–214 (2018).
35. Sorokina, I. T. & Sorokin, E. Femtosecond Cr2+-Based Lasers. *IEEE J. Sel. Top. Quantum Electron.* **21**, 273–291 (2015).
36. Mirov, S. B. *et al.* Frontiers of Mid-IR Lasers Based on Transition Metal Doped Chalcogenides. *IEEE J. Sel. Top. Quantum Electron.* **24**, 1–29 (2018).
37. Kowalczyk, M. *et al.* Ultra-CEP-stable single-cycle pulses at 2.2 µm. *Optica* **10**, 801–811 (2023).
38. Vasilyev, S. *et al.* Super-octave longwave mid-infrared coherent transients produced by optical rectification of few-cycle 2.5-µm pulses. *Optica* **6**, 111–114 (2019).
39. Steinleitner, P. *et al.* Single-cycle infrared waveform control. *Nat. Photonics* **16**, 512–518 (2022).
40. Heidrich, J. *et al.* Low-Noise Femtosecond SESAM Modelocked Diode-Pumped Cr:ZnS Oscillator. *IEEE J. Quantum Electron.* **59**, 1–7 (2023).
41. Vasilyev, S. *et al.* Ultra-Low Noise Cr:ZnS Laser Source for High Performance Dual Comb Spectroscopy. in *2024 Conference on Lasers and Electro-Optics (CLEO)* 1–2 (2024).
42. Giannotti, D. *et al.* Comprehensive intensity and phase noise analysis of a femtosecond Cr:ZnSe Kerr-lens mode-locked laser. *Opt. Laser Technol.* **184**, 112528 (2025).
43. Vasilyev, S. *et al.* Middle-IR frequency comb based on Cr:ZnS laser. *Opt. Express* **27**, 35079–35087 (2019).
44. Kim, D. *et al.* Intensity noise suppression in mode-locked fiber lasers by double optical bandpass filtering. *Opt. Lett.* **42**, 4095–4098 (2017).
45. McFerran, J. J., Swann, W. C., Washburn, B. R. & Newbury, N. R. Suppression of pump-induced frequency noise in fiber-laser frequency combs leading to sub-radian fceo phase excursions. *Appl. Phys. B* **86**, 219–227 (2007).
46. Wang, Y. *et al.* 47-fs Kerr-lens mode-locked Cr:ZnSe laser with high spectral purity. *Opt. Express* **25**, 25193–25200 (2017).
47. Barh, A. *et al.* High-power low-noise 2-GHz femtosecond laser oscillator at 2.4 µm. *Opt. Express* **30**, 5019–5025 (2022).
48. Hahner, D., Steinleitner, P., Chen, Y., Mak, K. F. & Pervak, V. Second and third-order dispersion compensating mirror pairs for the spectral range from 1.2-3.2µm. *Opt. Express* **30**, 38709–38716 (2022).
49. Sorokina, I. T. *et al.* Continuous-wave tunable Cr2+:ZnS laser. *Appl. Phys. B* **74**, 607–611 (2002).
50. Okazaki, D., Song, W., Morichika, I. & Ashihara, S. Mode-locked laser oscillation with spectral peaks at molecular rovibrational transition lines. *Opt. Lett.* **47**, 6077–6080 (2022).
51. Couairon, A. & Mysyrowicz, A. Femtosecond filamentation in transparent media. *Phys. Rep.* **441**, 47–189 (2007).
52. Borchers, B., Anderson, A. & Steinmeyer, G. On the role of shot noise in carrier-envelope phase stabilization: On the role of shot noise in CEP stabilization. *Laser Photonics Rev.* **8**, 303–315 (2014).
53. Okubo, S., Onae, A., Nakamura, K., Udem, T. & Inaba, H. Offset-free optical frequency comb self-referencing with an f-2f interferometer. *Optica* **5**, 188–192 (2018).
54. Kim, J., Kärtner, F. X. & Ludwig, F. Balanced optical-microwave phase detectors for optoelectronic phase-locked loops. *Opt. Lett.* **31**, 3659–3661 (2006).
55. Suliga, K., Sotor, J. & Kowalczyk, M. Direct electro-optic phase control for carrier-envelope offset frequency stabilization in solid-state lasers. *Opt. Express* **33**, 21870–21879 (2025).

56. Shoji, T. D. *et al.* Ultra-low-noise monolithic mode-locked solid-state laser. *Optica* **3**, 995–998 (2016).
57. Helbing, F. W., Steinmeyer, G., Stenger, J., Telle, H. R. & Keller, U. Carrier–envelope-offset dynamics and stabilization of femtosecond pulses. *Appl. Phys. B* **74**, s35–s42 (2002).
58. Kakehata, M. *et al.* Single-shot measurement of carrier-envelope phase changes by spectral interferometry. *Opt. Lett.* **26**, 1436–1438 (2001).
59. Qu, S. *et al.* Directly diode-pumped femtosecond Cr:ZnS amplifier with ultra-low intensity noise. *Opt. Lett.* **47**, 6217–6220 (2022).
60. Leshchenko, V. E. *et al.* High-power few-cycle Cr:ZnSe mid-infrared source for attosecond soft x-ray physics. *Optica* **7**, 981–988 (2020).
61. Danilin, R. *et al.* Single-Pass Amplification of Ultrashort Pulses in Cr:ZnSe Gain Element Pumped by Radiation of Q-switched 1645 nm Er:YAG Laser. in *CLEO 2025 (2025), paper JPS200_87* JPS200_87 (Optica Publishing Group, 2025). doi:10.1364/CLEO_AT.2025.JPS200_87.
62. Liang, W. *et al.* MHz-Rate Cr:ZnS Amplifier Providing 2.7 W of Pulses with Sub-10-fs Transform Limit at 2.3 μm. in *2025 Conference on Lasers and Electro-Optics Europe & European Quantum Electronics Conference (CLEO/Europe-EQEC)* 1–1 (2025). doi:10.1109/CLEO/Europe-EQEC65582.2025.11111064.
63. Feng, G. *et al.* 10-W sub-100-fs ultrafast Cr:ZnS/ZnSe MOPA system enabled by doping gradient engineering. *Infrared Phys. Technol.* **157**, 106654 (2026).
64. Muraviev, A., Konnov, D., Vasilyev, S. & Vodopyanov, K. L. Dual-frequency-comb UV spectroscopy with one million resolved comb lines. *Optica* **11**, 1486–1489 (2024).
65. Sorokin, E. *et al.* Atmospheric dispersion management in mid-IR mode-locked oscillators. *Opt. Express* **31**, 18790–18798 (2023).
66. Liehl, A., Fehrenbacher, D., Sulzer, P., Leitenstorfer, A. & Seletskiy, D. V. Ultrabroadband out-of-loop characterization of the carrier-envelope phase noise of an offset-free Er:fiber frequency comb. *Opt. Lett.* **42**, 2050–2053 (2017).